\documentclass[%
article,
nofootinbib,
 amsmath,amssymb,
]{elsarticle}

\usepackage{graphicx}
\usepackage{xcolor}
\usepackage{amsmath,amssymb}
\usepackage{bm}
\usepackage{hyperref}
\usepackage{url}
\usepackage{color}
\usepackage{fleqn}
\usepackage{dcolumn}

\def\bea{\begin{eqnarray}}
\def\eea{\end{eqnarray}}
\def\be{\begin{equation}}
\def\ee{\end{equation}}
\newcommand{\Sec}[1]{Sec.~\ref{#1}}
\newcommand{\Eq}[1]{Eq.~\eqref{#1}}

\begin{document}
\begin{frontmatter}

\title{Searching for Light in the Darkness: Bounds on ALP Dark Matter with the optical MUSE-Faint survey}

\author{Marco Regis\fnref{myfootnote}}%
\address{Dipartimento di Fisica, Universit\`{a} di Torino, via P. Giuria 1, I--10125 Torino, Italy}
\address{Istituto Nazionale di Fisica Nucleare, Sezione di Torino, via P. Giuria 1, I--10125 Torino, Italy}
\fntext[myfootnote]{marco.regis@unito.it}

\author{Marco Taoso\fnref{mtfootnote}}
\address{Istituto Nazionale di Fisica Nucleare, Sezione di Torino, via P. Giuria 1, I--10125 Torino, Italy}
\fntext[mtfootnote]{marco.taoso@to.infn.it}

\author{Daniel Vaz}
\address{Instituto de Astrofísica e Ciências do Espaço, Universidade do Porto, CAUP, Rua das Estrelas, PT4150-762 Porto, Portugal}
\address{Departamento de F{\'\i}sica e Astronomia, Faculdade de Ci{\^e}ncias, Universidade do Porto,
Rua do Campo Alegre 687, PT4169-007 Porto, Portugal}

\author{Jarle Brinchmann}
\address{Instituto de Astrofísica e Ciências do Espaço, Universidade do Porto, CAUP, Rua das Estrelas, PT4150-762 Porto, Portugal}
\address{Leiden Observatory, Leiden University, P.O.~Box~9513, 2300~RA~Leiden, The Netherlands}

\author{Sebastiaan L. Zoutendijk}
\address{Leiden Observatory, Leiden University, P.O.~Box~9513, 2300~RA~Leiden, The Netherlands}

\author{Nicolas F. Bouch\'e}
\address{Univ. Lyon, Univ. Lyon1, ENS de Lyon, CNRS, Centre de Recherche Astrophysique de Lyon UMR5574, 69230, Saint-Genis-Laval, France}

\author{Matthias Steinmetz}
\address{Leibniz-Institut f\"ur Astrophysik, Potsdam (AIP), An der Sternwarte 16, 14482 Potsdam, Germany}

\begin{abstract}
We use MUSE spectroscopic observations of the dwarf spheroidal galaxy Leo T between 470 and 935 nm to search for radiative decays of axion like particles (ALPs). Under the assumption that ALPs constitute the dark matter component of the Leo T halo, we derive bounds on the effective ALP-two-photon coupling. We improve existing limits by more than one order of magnitude in the ALP mass range 2.7-5.3 eV.
\end{abstract}



\end{frontmatter}

\section{\label{sec:intro} Introduction}

Axion-like particles (ALPs) are compelling cold dark matter candidates~\cite{Preskill:1982cy,Abbott:1982af,Dine:1982ah}. They are a generalization of the QCD axion, which was  originally introduced to solve the strong charge-parity (CP) problem~\cite{Peccei:1977hh,Peccei:1977ur, Weinberg:1977ma, Wilczek:1977pj}. Numerous experimental strategies have been envisaged to detect ALPs (see Ref.~\cite{Irastorza:2018dyq} for a review), many of them exploiting their coupling to photons given by the operator $\mathcal{L}=-\frac{1}{4}g_{a\gamma\gamma}\,a\,F_{\mu\nu}\tilde{F}_{\mu\nu},$ where $a$ is the ALP field, $F_{\mu\nu}$ is the electromagnetic field strength, $\tilde{F}_{\mu\nu}$ its dual, and $g_{a\gamma\gamma}$ the coupling constant.
In astrophysical environments, this operator leads to photon signals from the radiative decay of ALP dark matter~\cite{Caputo:2018vmy,Caputo:2018ljp,Kephart:1994uy,Rosa:2017ury,Tkachev:1987cd,Battye:2019aco,Sigl:2019pmj} 
and the conversion of ALPs into photons in the presence of  magnetic fields~\cite{Hook:2018iia,Huang:2018lxq,Safdi:2018oeu,Edwards:2019tzf,Leroy:2019ghm,Dessert:2019sgw,Sigl:2017sew}.
For ALP masses in the eV range, the monochromatic line emission from ALP decays falls in the optical and near infrared bands.
Upper limits on this signal have been obtained from observations of galaxy clusters in Ref.~\cite{Grin:2006aw}, see also Refs.~\cite{Ressell:1991zv,Bershady:1990sw} for previous analysis. Interestingly, ALPs of such masses have been proposed to explain an excess of the measured cosmic near infrared background and its angular anisotropies~\cite{Gong:2015hke,Kalashev:2018bra}.
Recently the XENON1T experiment has reported an excess of electron recoil events over expected backgrounds with a significance around $3.5\sigma$~\cite{Aprile:2020tmw}. This anomaly can be interpreted in terms of ALPs with masses $<100$ eV produced in the interior of the Sun and then detected by the Xenon1T experiment through the axioelectric effect, exploiting the couplings of ALPs with electrons~\footnote{Notice however that this interpretation is in tension with astrophysical constraints, see~\cite{Aprile:2020tmw}}.
While this result calls for further examination, it certainly spurs interest in ALPs.

Dwarf spheroidal galaxies are ideal targets to search for indirect dark matter signals, given their proximity and their high dark matter content. In this work, assuming that all dark matter is in the form of ALPs, we constrain its coupling to photons by searching for the optical line from ALP decays in the Leo T dwarf spheroidal galaxy. We exploit data from a Guaranteed Time Observing (GTO) programme targeting ultra-faint dwarf galaxies using the Multi Unit Spectroscopic Explorer (MUSE) at the VLT \cite{MUSE2017}.  These spectroscopic data cover the range 470-935 nm with a medium spectra resolution ($R=\lambda/\Delta \lambda>10^3$) and excellent sensitivity, covering the area around the centre of Leo T, extending up to approximately its half-light radius.

Previous astrophysical studies \citep{Grin:2006aw} placed an upper limit on $g_{\alpha \gamma \gamma}$ of  $\approx 5\times 10^{-12}\,\mathrm{GeV}^{-1}$ for ALP masses between 4.5 and 5.5 eV. In the following we will demonstrate that the MUSE data on Leo T  improves current constraints by more of an order of magnitude for ALP masses between 2.7 and 5.3 eV.

The paper is structured as follows. The data from MUSE observations are presented in \Sec{sec:data}. The computation of the ALP signal is detailed in \Sec{sec:axion}. The statistical analysis and results are discussed in~\Sec{sec:res}. We conclude in \Sec{sec:conc}.

\section{\label{sec:data} Observations and data reduction}

The central region of Leo T was observed as part of MUSE-Faint \citep{2020A&A...635A.107Z}, a GTO survey of ultra-faint dwarf galaxies (PI Brinchmann) with MUSEww, a large-field medium-resolution Integral Field Spectrograph on the Very Large Telescope (VLT). The data described here use 15 exposures of 900 seconds adding up to 3.75 hours of exposure time\footnote{Run IDs 0100.D-0807, 0101.D-0300, 0102.D-0372 and 0103.D-0705}, see also Vaz et al (in preparation). The data were taken in the Wide Field Mode with adaptive optics (WFM-AO), which provides a $1 \times 1 \; \mathrm{arcmin^2}$ field of view with a spatial sampling of  $0.2\;\mathrm{arcsec\;pixel^{-1}}$ and with a spatial resolution of $0.61\;\mathrm{arcsec}$ (full width-half maximum) at a wavelength of $7000$ \AA. The data covers a wavelength range of $4700-9350$ \AA\ with a wavelength sampling of $1.25$\AA. In order to avoid the light of the sodium laser of the adaptive optics system, a blocking filter removes light in the wavelength range 5820-5970\AA\ (2.13--2.08 eV) which shows up as a gap in the constraints below.

Standard data reduction using the MUSE Data Reduction Software (DRS; version 2.8.1; Ref.~\cite{2020arXiv200608638W}) was used, following closely the methodology described in \cite{2020A&A...635A.107Z}. The most salient points for this paper being that the data were flux calibrated using flux standards observed during the night, atmospheric emission lines, caused by Raman scattering of the laser light of the adaptive optics system, were removed, and we performed a subtraction of emission lines from the night sky. These latter have well-known wavelengths and lead to increased noise at these wavelengths.

Since a key aim of this paper is to place limits on the presence of emission lines across the field of view of Leo T, it is important that we have a good estimate of the noise in the data cube. The standard noise estimate is based on the uncertainty propagated through the DRS. The photon count is high enough that its uncertainty can be considered Gaussian. Moreover the individual steps of the DRS are linear, and treat each pixel independently. However, it is known \citep[e.g.][]{2017A&A...608A...1B} that the MUSE DRS underestimates the uncertainties in the final data-cube since covariance terms are neglected in the final interpolation step. To address this point we re-estimated the pixel-to-pixel variance directly from each individual datacube following the approach described in Ref.~\cite{2017A&A...608A...1B}, using SExtractor~\cite{1996A&AS..117..393B} to define a mask image. Finally, all single exposure datacubes were combined using MPDAF~\cite{2017arXiv171003554P}, creating the datacube that has been used in the following part of the analysis. We discuss the impact of different noise estimates have on the results further below.

The data contain a large number of stellar sources within the field of view, many from Leo T (to be discussed in Vaz et al in prep), but also some likely foreground stars from the Milky way. Moreover also some galaxies are present. 
To reduce the impact of these sources on the final results we mask the brightest ones.  
To construct the mask we took two steps: first we created a white-light image by summing over the wavelength axis, which increases the signal-to-noise in each source ensuring that even sources that are undetected in a single wavelength layer are masked out. We then ran SExtractor on the white-light image, with a detection threshold of $3\sigma$ (the results are not very sensitive to this particular choice, see also below). The resulting segmentation map is used to mask sources, while no attempt to subtract them has been made. Thus we will mainly consider data where no sources are detected in the white-light image.

Let us point out two caveats to this analysis.
(a) Emission from the night sky cannot be perfectly subtracted, which has two consequences: one is that there are residuals, both positive and negative, around the strongest sky emission lines and the other is that the mean flux where no objects are found may differ from zero after sky subtraction.
(b) There are a number of galaxies in the data cube that are mostly seen as line emitters and not identified as sources with the above procedure. Note that faint stars have a smooth spectrum with energy interspersed with absorption features and should not confound our results. 

For (a) we could apply a sky subtraction correction code \citep[e.g.\ ZAP][]{2016MNRAS.458.3210S}, but this procedure might add subtle effects in the data that are undesirable for our analysis. Instead we 
include an arbitrary spatially flat term in each wavelength layer (and set it through the fit) to account for any zero-point offset. To address b) we used the ORIGIN emission line detection software \citep{2020A&A...635A.194M} to blindly detect emission line sources which we then mask out. These emission line sources turn out to be rare and their masking does not change the results significantly.

\begin{figure}[ht!]
\vspace{-15mm}
   \includegraphics[width=0.53\textwidth]{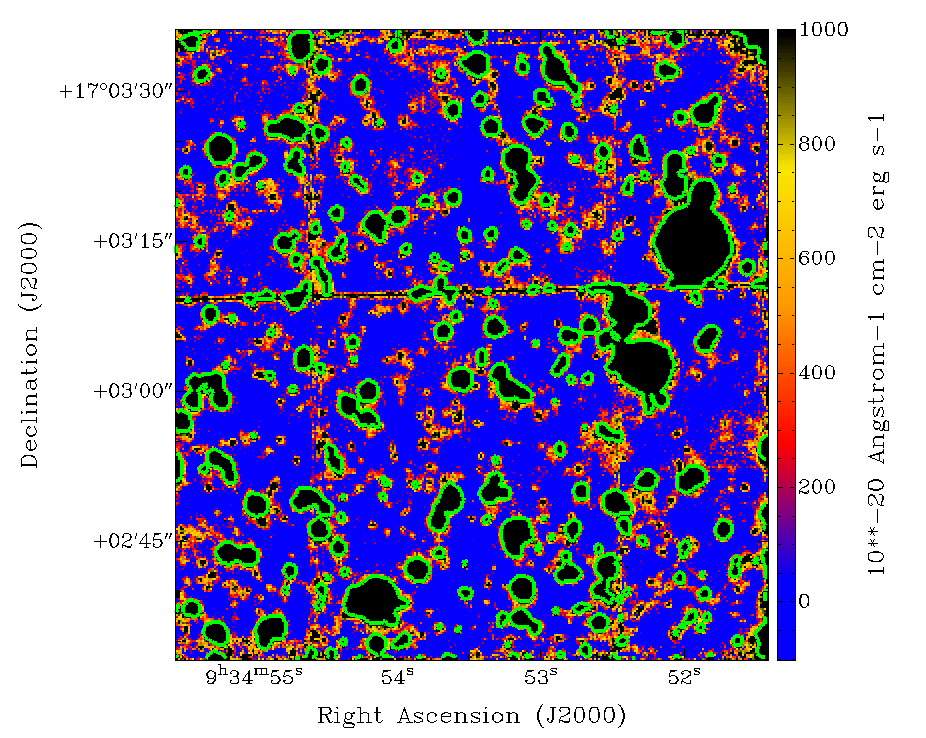}
   \includegraphics[width=0.45\textwidth]{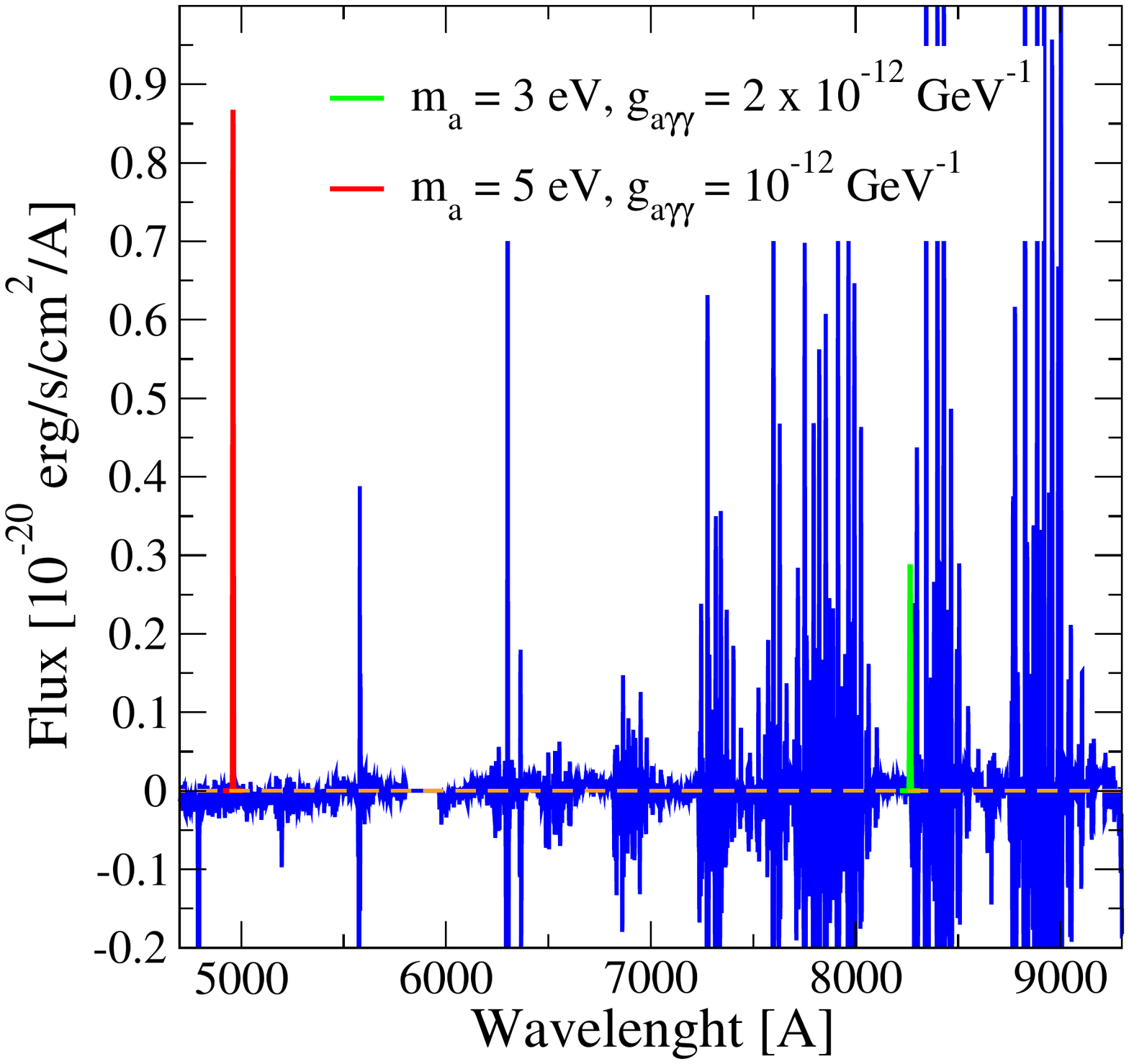}
    \caption{{\bf Left panel:} White-light image of Leo T area obtained with MUSE observations. Contours from segmentation map obtained running SExtractor are overlaid in green, see text for more details.
{\bf Right panel:} Average flux density as a function of the wavelength of observation. For illustrative purposes, we include the expected signal from an ALP with $g_{a\gamma\gamma}=2\times 10^{-12}\,{\rm GeV}^{-1}$ and mass $m_a=3$ eV (green) and an ALP with $g_{a\gamma\gamma}=10^{-12}\,{\rm GeV}^{-1}$ and $m_a=5$ eV (red).}
\label{fig:illustr}
 \end{figure}

In the left panel of Fig.~\ref{fig:illustr}, we show the aforementioned white-light image of Leo T with the 3$\sigma$ segmentation mask overlaid. In the right panel we report with a blue line the flux density averaged over all the unmasked pixels of our region of interest and in the different wavelength channels of MUSE observations.

\section{\label{sec:axion} ALP signal}

The flux density at wavelength $\lambda$ produced by decays of ALPs inside Leo T from a given direction identified by $\theta$ can be computed as:
\be
S_\lambda (\theta)=\frac{\Gamma_a}{4\pi}\,\frac{1}{\sqrt{2\pi}\sigma_\lambda} \exp{\left[-\frac{(\lambda-\lambda_{em})^2}{2\sigma_\lambda^2}\right]}\int d\Omega\, d\ell \rho_a[r(\theta,\Omega,\ell)]\,B(\Omega)\;,
\label{eq:flux}
\ee
where we have assumed spherical symmetry. Deviations from spherical symmetry are constrained to small values for LeoT~\cite{Hayashi:2016kcy}, and, in any case, their possible presence would have a minor impact on the final bounds.

The decay rate $\Gamma_a$ depends on the the ALP mass $m_a$ and the effective ALP-two-photon coupling $g_{a\gamma\gamma}$. In natural units, it reads
$\Gamma_a=m_a^3 \,g_{a\gamma\gamma}^2/(64\pi)$.

The dark matter spatial density distribution $\rho_a(r)$ is written as a function of the distance $r$ from the center of the dwarf.
This distance can be expressed in terms of the coordinate along the line of sight $\ell$, the angle $\theta^\prime$ and the distance $D$ of Leo T from us by means of $r^2=\ell^2+D^2-2\,\ell\,D\cos\theta^\prime.$ The angle $\theta^\prime$ combines the direction of observation $\theta$, i.e. the angular off-set with respect to the
Leo T centre, and the position $\Omega$ inside the observing angular beam.
The latter is given by the function $B(\Omega)$.
We assume a Gaussian function for both the energy and angular response of the detector.  Their FWHM as a function of wavelength are taken from Ref.~\cite{2017A&A608A1B}, and vary between 0.75 arcsec to 0.55 arcsec
in the wavelength range of 4700-9350 \AA.
In \Eq{eq:flux} $\sigma_\lambda$ denotes the spectral resolution.

The wavelength of emission can be computed as $\lambda_{em}=c/\nu_{em}$ with $\nu_{em}=m_a/(4\pi)$. We neglect the velocity dispersion of ALPs in the Leo T
halo, since it is significantly smaller than the spectral resolution of MUSE.
Indeed, the velocity dispersion is $\sigma_v\lesssim 10$ km/s~\cite{Bonnivard:2015xpq}, which means $\sigma_v/c\lesssim 3\times 10^{-5}$, while the spectral resolution is $\Delta\lambda/\lambda\gtrsim 5\times 10^{-4}$.

In our analysis we scan over $g_{a\gamma\gamma}$ and $m_a$, while we take a model for the ALP spatial distribution. Analyses of the velocity dispersion in dSph often provides the so-called $D$-factor. 
It is defined as:
\be
D(\tilde \theta_{max})=\int_{\Delta\Omega} d\ell\, d\Omega \, \rho_{DM}[r(\ell,\Omega)] =\int_{\Delta\Omega} d\Omega \,\phi(\Omega)
\label{eq:Dfactor}
\ee
where $d\Omega=2\pi\sin\tilde \theta d\tilde \theta$ and the angular region of integration is identified by the angle $\tilde \theta_{max}$.
For our computation, we are interested in the function $\phi=\int d\ell \rho_{DM}$, which enters in Eq.~\ref{eq:flux}.
To obtain it we consider the function $D(\theta)$ determined in Ref.~\cite{Bonnivard:2015xpq} from observational data, and then we invert Eq.~\ref{eq:Dfactor}.
The uncertainties we will quote in 
\Sec{sec:res} are derived from the uncertainty on the $D$-factor from Ref.~\cite{Bonnivard:2015xpq}. The latter is about a factor of four (at 68\% C.L.) at $\sim 1'$, which translates in approximately a factor of two for what concerns the bound on $g_{a\gamma\gamma}$.

The flux density in Eq.1 grows $\propto g_{a\gamma\gamma}^2$ and depends on the ALP mass through the ALP decay rate ($\Gamma_a\propto m_a^3$), the spectral resolution and the observing angular beam at the wavelength of emission.
In Fig.~\ref{fig:illustr} (right), we show the expected signal in Leo T, taking two examples of ALPs with couplings significantly smaller than any existing bound: $g_{a\gamma\gamma}=2\times 10^{-12}\,{\rm GeV}^{-1}$ for $m_a=3$ eV (green) and $g_{a\gamma\gamma}=10^{-12}\,{\rm GeV}^{-1}$ for $m_a=5$ eV (red).
Comparing them with the measured average emission (blue line), it is clear that such models are in the ball-park accessible by MUSE observations.

\section{\label{sec:res} Statistical Analysis and Results}
We assume the likelihood for the ALP diffuse emission to be described by a Gaussian likelihood:
\be 
\mathcal{L}=e^{-\chi^2/2} \;\;\; {\rm with} \;\;\; \chi^2=\frac{1}{N_{pix}^{FWHM}}\sum_{i=1}^{N_{pix}} \left(\frac{S_{th}^i-S_{obs}^i}{\sigma_{rms}^i}\right)^2\;,
\label{eq:like}
\ee
where $S_{th}^i$ is the theoretical estimate for the flux density in the pixel $i$, $S_{obs}^i$ is the observed flux density and $\sigma_{rms}^i$ is the r.m.s. error, both described in \Sec{sec:data}. 
The theoretical estimate is provided by Eq.~\ref{eq:flux} plus a spatially flat term $S_{\lambda,flat}$ that we include in the fit to the map at each wavelength (to account for incomplete sky subtraction) and we treat as a nuisance parameter.
$N_{pix}$ is the total number of pixels (around $7\times 10^4$, with very small dependence on frequency) in the area under investigations, 
that is chosen to be a circle of $30''$ of radius. $N_{pix}^{FWHM}$ is the number of pixels within the MUSE angular FWHM.

Bounds on the parameter $g_{a\gamma\gamma}$ are computed at any given mass $m_a$
through a profile likelihood technique, namely ``profiling out'' the nuisance parameter $S_{\lambda,flat}$.
We assume that $\lambda_c(g_{a\gamma\gamma})=-2\ln[\mathcal{L}(g_{a\gamma\gamma},S_{\lambda,flat}^{b.f.})/\mathcal{L}(g_{a\gamma\gamma}^{b.f.},S_{\lambda,flat}^{b.f.})]$ follows a $\chi^2$-distribution with one d.o.f. and with one-sided probability given by $P=\int^{\infty}_{\sqrt{\lambda_c}}d\chi\,e^{-\chi^2/2}/\sqrt{2\,\pi}$, where $g_{a\gamma\gamma}^{b.f.}$ denotes the best-fit value for the coupling at that specific ALP mass.
In other words, the 95\% C.L. upper limit on $g_{a\gamma\gamma}$ at mass $m_a$ is obtained by increasing the signal from its best-fit value until $\lambda_c=2.71$, keeping $S_{\lambda,flat}$ fixed to its best-fit value.

\begin{figure}[ht!]
\vspace{-25mm}
\centering
   \includegraphics[width=0.7\textwidth]{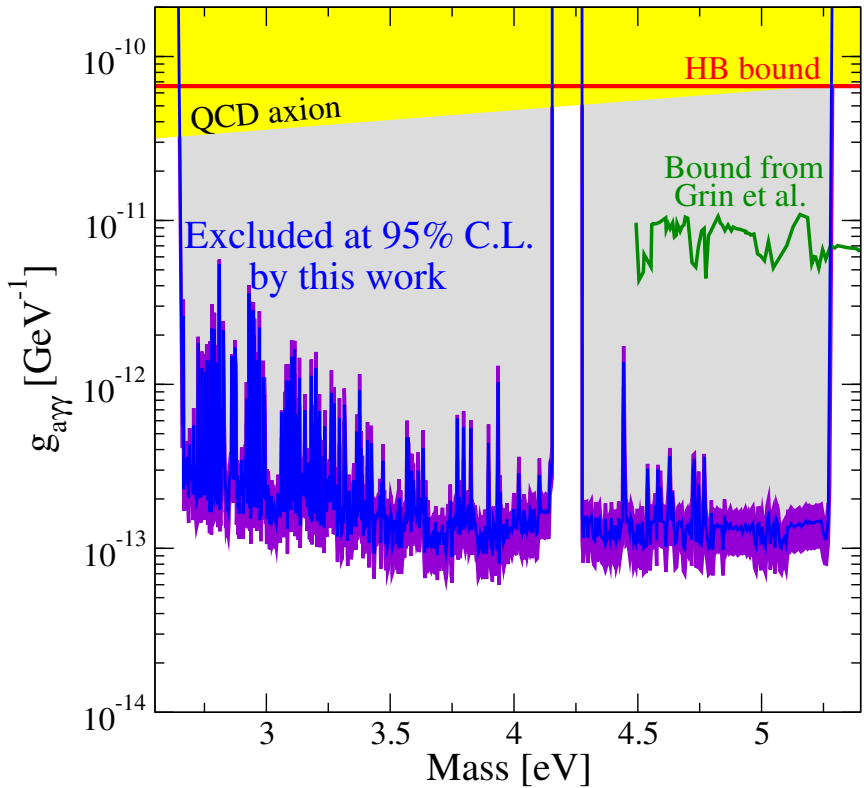}
    \caption{The solid blue curve shows the 95\% C.L. upper limits on the effective ALP-two-photon coupling $g_{a\gamma\gamma}$ as a function of the ALP mass, derived in this work.
    The violet area includes the uncertainties on the $D$-factor, taken from Ref.~\cite{Bonnivard:2015xpq}.
    The gap in the constraints is due to a blocking filter, used to remove the light of the sodium laser of the adaptive optics system.
    We also show the bound derived in Ref.~\cite{Grin:2006aw} from observation of clusters, in Ref.~\cite{Ayala:2014pea} from the ratio of horizontal branch (HB) to red giants stars in globular clusters,  and the preferred region for the QCD axion~\cite{DiLuzio:2016sbl}}
\label{fig:bounds}
 \end{figure}

Results are shown in Fig.~\ref{fig:bounds}.
The solid blue curve refers to the bounds derived assuming the best-fit $D$-factor from Ref.~\cite{Bonnivard:2015xpq}, while the shaded area considers their 68\% C.L. interval for the $D$-factor.
The curves show rapid variation at lower energies/longer wavelengths reflecting the presence of strong OH emission lines from the night sky which increases the noise here, as mentioned in Section~\ref{sec:data}.
The limit on $g_{a\gamma\gamma}$ becomes more stringent as the mass $m_a$ increases. This is due to the scaling of the decay rate as $\Gamma_a \propto m_a^3$ (mitigated by the fact that also the background increases as the wavelength decreases).
Another estimate of the D-factor can be obtained using the results of ref.\,\cite{Faerman:2013pmm}, also considered in ref. \cite{Wadekar:2019xnf}, which exploits 21 cm radio observations to infer the dark matter distribution of Leo T, described with NFW and Burkert profiles. We have found that the constraints on $g_{a\gamma\gamma}$ based on those profiles are slightly less stringent than our reference analysis, but typically fall inside the uncertainty band in fig.2.

In Fig.~\ref{fig:bounds}, we also show the bound derived in Ref.~\cite{Grin:2006aw} from observation of clusters, in Ref.~\cite{Ayala:2014pea} from the ratio of horizontal branch (HB) to red giants stars in globular clusters and, for reference, the preferred region for the QCD axion~\cite{DiLuzio:2016sbl} \footnote{Notice however than for the masses under consideration, the QCD axion is excluded by astrophysical and laboratory probes, associated to couplings different from $g_{a\gamma\gamma}$, see e.g.~\cite{Tanabashi:2018oca}}. 
Our results improve existing bounds by more than one order of magnitude. 
They exclude the possible interpretation of near infrared
background anisotropies in terms of ALP dark matter~\cite{Gong:2015hke} in the wavelength/mass range covered by our analysis.

To test the robustness of our results against different masking and error estimates, we conduct a few sanity checks.

\begin{figure}[ht!]
\vspace{-15mm}
   \includegraphics[width=0.48\textwidth]{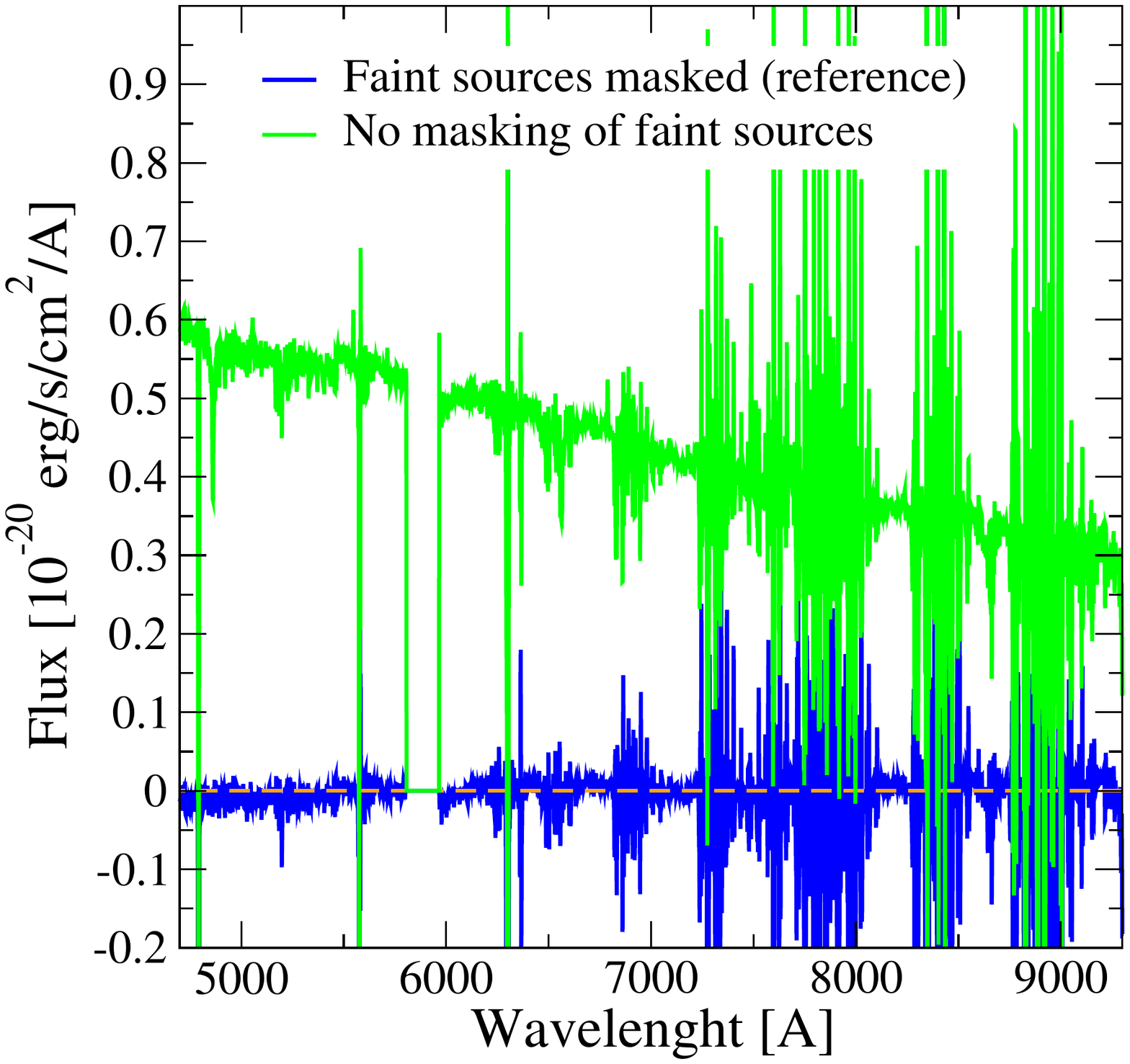}
\hspace{5mm}
   \includegraphics[width=0.46\textwidth]{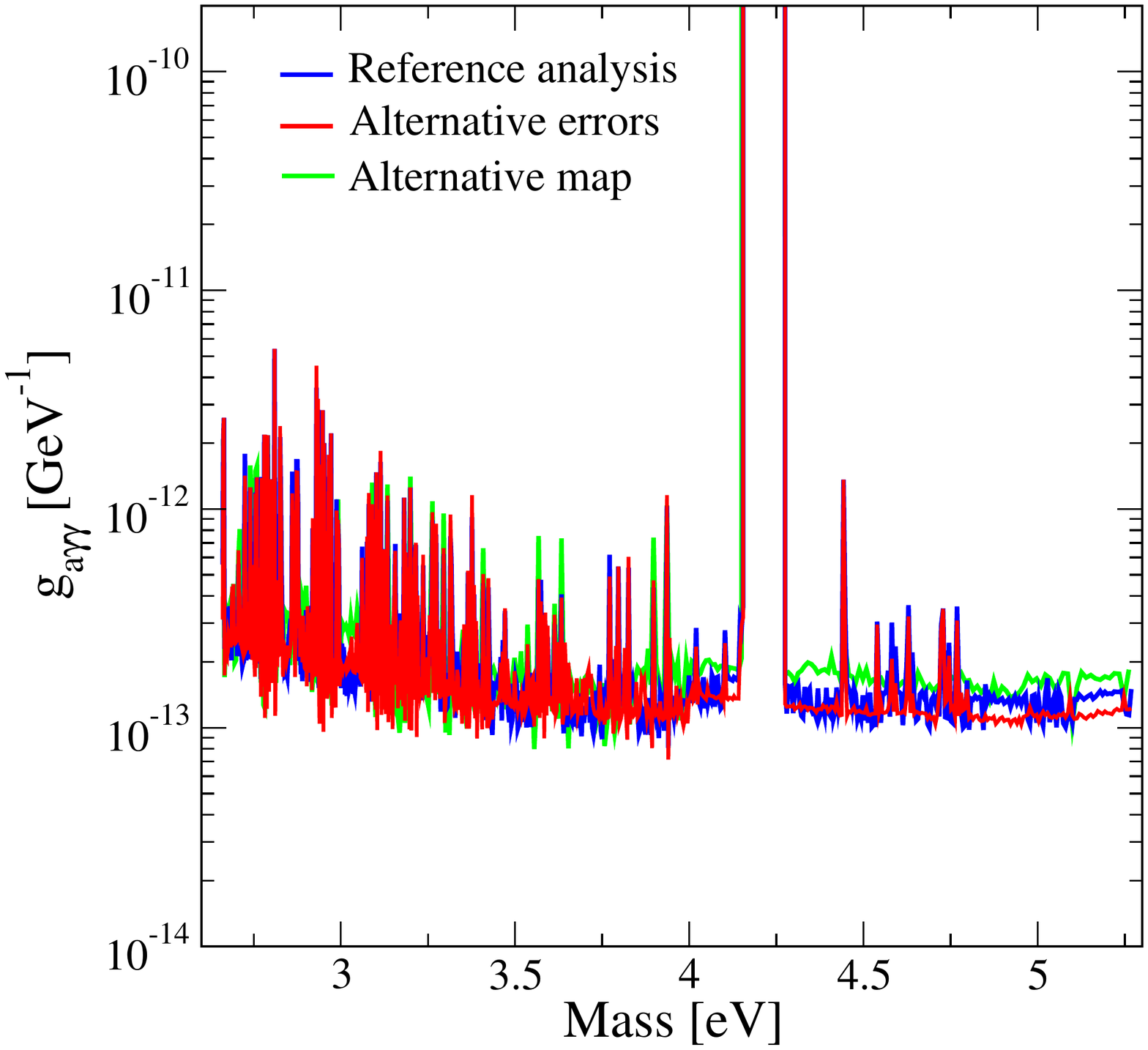}
    \caption{{\bf Left panel:} Comparison of the average measured spectrum obtained with/without masking faint sources. 
{\bf Right panel:} 95\% C.L. upper limits on the effective ALP-two-photon coupling $g_{a\gamma\gamma}$ as a function of the ALP mass, derived with alternative derivation of the errors (red) and without masking faint sources (green), compared with the reference analysis (blue). See text for details.}
\label{fig:app}
 \end{figure}

First, we perform the same analysis mentioned above but on maps where we discard the last step of data manipulation described in Section~\ref{sec:data}.
Namely, we do not mask the faint sources (which are derived through segmentation of the white-light image by SExtractor and by running ORIGIN for emission line sources). The spatially averaged spectrum of these maps is shown in green in Fig.~\ref{fig:app} (left). The spectrum is well above zero at nearly any wavelength, meaning that there is a significant residual emission, not related to the ALP signal. On the other hand, the resulting bounds (green line in the right panel) are only very mildly less constraining than in the reference analysis (blue line). At few wavelengths, the bounds becomes slightly more constraining in the unmasked case because in the masked map the fit shows a preference for the dark matter component over the flat term.

As a second test, we consider a different derivation of the measurement uncertainties, by computing the standard deviation in a region of $2''\times 2''$ around the pixel $i$, instead of using the procedure described in Section~\ref{sec:data}. We find this alternative derivation to provide, on average, a slightly more optimistic estimate of the errors with respect to the reference analysis.
However, the bounds are only marginally different, as can be seen by comparing red and blue lines in the right panel of Fig.~\ref{fig:app}, supporting the reliability of our analysis.

Finally, we have also tested that varying the spectral resolution within its uncertainties has a negligible impact on the bounds.

\section{\label{sec:conc} Conclusions}
Dark matter in the form of ALPs can be searched for exploiting their coupling to photons. One strategy is to look for the monochromatic photon flux generated by ALP decays inside astrophysical structures. Dwarf spheroidal galaxies are particularly suitable targets since they are dark matter dominated objects and they are relatively close to us. 
Excellent spectral resolution and sensitivity are required to search for the faint lines from ALPs.
For this purpose, we have exploited the spectroscopic observations obtained with the MUSE instrument at the VLT.
In this work we have searched for ALPs signals from the Leo T dwarf spheroidal galaxy.
The data cover wavelengths in the range 470-935 nm and therefore allow to test ALP masses between 2.7 and 5.3 eV.

We have derived new bounds on the effective ALP-two-photon couplings, which improve existing constraints from observations of clusters~\cite{Grin:2006aw} and stars in globular clusters~\cite{Ayala:2014pea} by more than one order of magnitude. 
We have investigated the impact of different sources of systematical uncertainties in the analysis, namely the masking of faint sources present in our image, the uncertainties on the measurements, and those on the spectral resolution. In this respect, we have shown that our constraints are rather robust.

In the future we plan to extend our analysis including MUSE observations of additional dwarf spheroidal galaxies. A joint analysis of several targets will allow not only to improve current exclusion limits, but it will also provide a better handle to unveil a possible ALP signal, which should obviously show up at the same wavelength in all the targets.

\bigskip \bigskip\bigskip
We would like to warmly thank Eline Tolstoy for her advice in the initial phase of the project.

MT acknowledge support from the INFN grant ‘LINDARK’ and the research grant ‘The Dark Universe: A Synergic Multimessenger Approach’ No. 2017X7X85K funded by MIUR.
MR and MT acknowledge support from the project ``Theoretical Astroparticle Physics (TAsP)'' funded by the INFN.

MR acknowledges support from: `Departments of Excellence 2018-2022' grant awarded by the Italian Ministry of Education, University and Research (\textsc{miur}) L.\ 232/2016; Research grant `From Darklight to Dark Matter: understanding the galaxy/matter  connection to measure the Universe' No.\ 20179P3PKJ funded by \textsc{miur}; Research grant `Deciphering the high-energy sky via cross correlation' funded by the agreement ASI-INAF n.2017-14-H.0. 
JB and DV acnowledge support by Fundação para a Ciência e a Tecnologia (FCT) through the
research grants UID/FIS/04434/2019, UIDB/04434/2020, UIDP/04434/2020 and JB also acknowledges support through the Investigador FCT Contract No. IF/01654/2014/CP1215/CT0003.
SLZ acknowledges support by The Netherlands Organisation for Scientific Research (NWO) through a TOP Grant Module 1 under project number 614.001.652.

Based on observations made with ESO Telescopes at the La Silla Paranal Observatory under programme IDs 0100.D-0807, 0101.D-0300, 0102.D-0372 and 0103.D-0705.


\bibliography{biblio}

\end{document}